\documentclass[%
 twocolumn,
 reprint,
showpacs,
 amsmath,amssymb,
 aps,
]{revtex4}

\usepackage{graphicx}
\usepackage{dcolumn}
\usepackage{bm}

\begin{document}

\preprint{APS/123-QED}

\title{Quantum Monte Carlo method for pairing phenomena:\\Super-counter-fluid of two-species Bose gases in optical lattices}

\author{Takahiro Ohgoe}
\author{Naoki Kawashima}%
\affiliation{%
	Institute for Solid State Physics, University of Tokyo, 5-1-5 Kashiwa-no-ha, Kashiwa, Chiba 277-8581, Japan
}%

\date{\today}

\begin{abstract}
We study the super-counter-fluid(SCF) states in the two-component hardcore Bose-Hubbard model on the square lattice, using the quantum Monte Carlo method based on the worm(directed loop) algorithm. Since the SCF state is a state of a pair-condensation characterized by $\langle a^{\dagger} b \rangle \neq 0, \langle a \rangle = 0$, and $\langle b \rangle = 0$, where $a$ and $b$ are the order parameters of the two components, it is important to study behaviors of the pair-correlation function $\langle a_{i} b_{i}^{\dagger} a_{j}^{\dagger} b_{j} \rangle$. For this purpose, we propose a choice of the worm head for calculating the pair-correlation function. From this pair-correlation, we confirm the Kosterlitz-Thouless(KT) charactor of the SCF phase. The simulation efficiency is also improved in the SCF phase.     

\end{abstract}

\pacs{02.70.Ss, 02.70.Tt, 03.75.Hh, 05.30.Jp}
\maketitle


	Since the pioneering work by Greiner {\it et al}.\cite{greiner2002}, there have been experimental developments in the field of ultra cold atoms in optical lattices. On theoretical sides, the Bose-Hubbard model which is the effective model of such systems\cite{jaksch1998} have been well studied. The exact ground state phase diagrams of the standard Bose-Hubbard model have been obtained by the quantum Monte Carlo method\cite{capogrosso2007, capogrosso2008, kawashima2009}. Recently, multi-component systems have been realized experimentally\cite{ospelkaus2006, gunter2006, catani2008}. For example, Catani {\it et al}.\cite{catani2008} trapped heteronuclear bosonic mixtures of $^{87}$Rb-$^{41}$K in optical lattices. Due to the multi-degree of freedom, several exotic phases have been predicted theoretically in this system\cite{soyler2009}. One of such phases is the super-counter-fluid(SCF) phase\cite{kuklov2003}. This state is a pair-condensation where A-particles and B-holes(A and B represent each component) are paired by the strong repulsive interspecies interaction. The SCF state is characterized by $\langle a^{\dagger} b \rangle \neq 0, \langle a \rangle = 0$, and $\langle b \rangle = 0$. In Ref \cite{soyler2009}, the ground state phase diagram of two-component hardcore Bose-Hubbard model at a commensurate filling is revealed by the quantum Monte Carlo simulations based on the worm algorithm\cite{prokofev1998}. The phase diagram consists of a number of distinct phases and the SCF is one of them that appears in the strongly correlated region when the asymmetry between the two kinds of particles is weak.

As mentioned above, the quantum Monte Carlo method is a powerful tool for investigating quantum many-body systems. The worm(directed loop) algorithm\cite{prokofev1998, syljuasen2002} is one of the most effective quantum Monte Carlo methods. In this algorithm, we can measure off-diagonal quantities such as two-point correlation function $\langle a_{i} a_{j}^{\dagger} \rangle$. However, in the simulation of two-species particle systems, the pairing correlation function $\langle a_{i} b_{i}^{\dagger} a_{j}^{\dagger} b_{j} \rangle$ can not be measured by the conventional method. Instead, several kinds of stiffnesses have often been used to detect the SCF state\cite{soyler2009, kuklov2004}. In order to measure the pair-correlation function which is related to the order parameter more directly, we can use the quantum Monte Carlo method proposed in Refs\cite{rombouts2006, houcke2006}. In the method, we sample the canonical ensembles by updates based on a worm operator which is always local in the imaginary time. This leads to much better statistics for equal-time off-diagonal quantities including the pair-correlation function. Although this method needs unconventional update procedures, this is valid when we simulate the system of fixed particle numbers. Since we can work in the canonical ensemble, we do not need to adjust chemical potentials to obtain desired fillings.

	In this Letter, we present a simple technique of measuring the pair-correlation function by the quantum Monte Carlo method in the grand canonical ensemble. In this method, we can perform conventional update procedures which needs only slightly extension. By this method, we study properties of the novel SCF state. Since the SCF phase is characterized by the order parameter $\langle a^{\dagger} b \rangle$, we discuss behaviors of the pair-correlation function $\langle a_{i} b_{i}^{\dagger} a_{j}^{\dagger} b_{j} \rangle$. Furthermore, we demonstrate this kind of worm improves simulation efficiency.
	
	\begin{figure*}[ht] 
		\includegraphics[width=15.7cm]{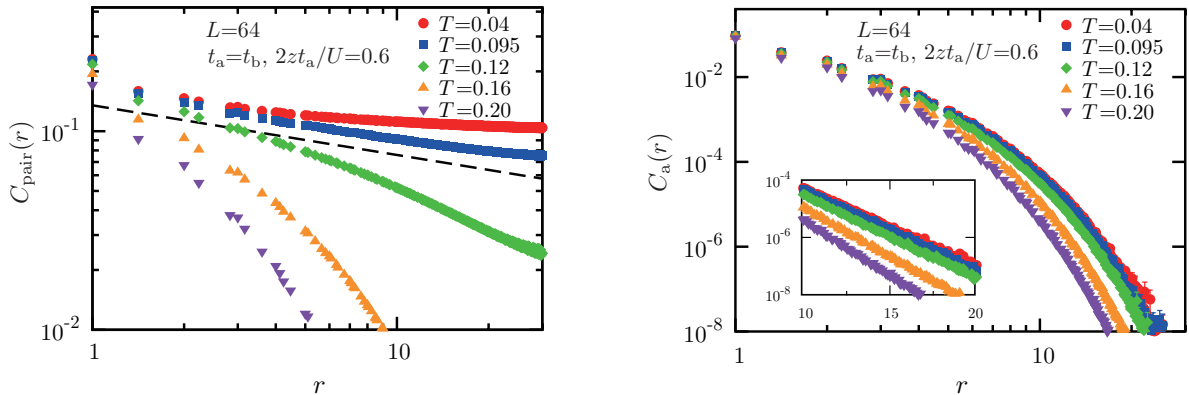}
		\caption{\label{fig:correl} (Color online) Logarithmic plots of $C_{\rm pair}(r)$(left panel) and $C_{\rm a}(r)$(right panel) at different temperatures. In the left panel, the dashed line represents $r^{1/4}$ which is expected at the critical temperature in the KT transition. In the right panel, the inset is semi-logarithmic plots of the large $r$ region where exponential decays can be observed. Error bars are drawn but most of them are smaller than the symbol size(here and the following figures). We take $t_{\rm a}$ as units of temperature.}
	\end{figure*}

		
	The model we considered here is the two-component hardcore Bose-Hubbard model on a squared lattice. The Hamiltonian is given by
	\begin{eqnarray}
			H & = & - \sum_{\langle i, j \rangle} t_{\rm a} a^{\dagger}_{i} a_{j} - \sum_{\langle i, j \rangle} t_{\rm b} b^{\dagger}_{i} b_{j}  + \sum_{i} U n_{{\rm a} i} n_{{\rm b} i},
	\end{eqnarray}
	where $a^{\dagger}_{i}( a_{i} )$ and $b^{\dagger}_{i}( b_{i} )$ are the bosonic creation(annihilation) operators on the site $i$ for two species of boson(A and B), $n_{{\rm a} i}=a^{\dagger}_{i} a_{i}$, $t_{\rm a}$($t_{\rm b}$) is the hopping parameter of A(B)-bosons, and $U$($>$0) represents the repulsive interspecies interactions. Each species are hardcore bosons, which means that two bosons of the same kind cannot occupy the same site. The summation $\langle i, j \rangle$ is over the nearest-neighbor pairs and the system size is defined by $N=L^{2}$. The periodic boundary condition is applied. In what follows, we consider the case of half-filling for each component, $\langle n_{{\rm a} i} \rangle = \langle n_{{\rm b} i} \rangle = 1/2$, where $\langle \ \rangle$ means the thermal average.

	We used the quantum Monte Carlo method based on the directed loop algorithm(DLA)\cite{syljuasen2002, kato2009}. In this algorithm, $d$-dimensional quantum systems are mapped to ($d+1$)-dimensional classical systems, using the Feynman path integral representation. The classical systems are called world-lines, because these ($d+1$)-dimensional systems are composed of the $d$-dimensional classical systems and the imaginary time axis. In the world-line quantum Monte Carlo method, world-line configurations are sampled. 
								
	The most characteristic feature of the DLA is update procedures. For updates, we create two discontinuous points called {\it a worm} in world-lines. In order to define weights of configurations which contain worms, we consider the Hamiltonian $H - \eta_{\rm a} Q_{\rm a} - \eta_{\rm b} Q_{\rm b}$, where the source term $Q_{\rm a}$ is given by 
	\begin{eqnarray}
		 Q_{\rm a}=\sum_{i} \int_{0}^{\beta} d \tau \{ a_{i}^{\dagger} (\tau ) + a_{i} (\tau ) \} /2,
	\end{eqnarray}
	and $Q_{\rm b}$ is defined likewise. In the above equation, $\tau$ is the imaginary time, $\beta$ represents the inverse temperature, and $a_{i}(\tau)= e^{\tau H} a_{i} e^{-\tau H}$. $\eta_{\rm a}$ and $\eta_{\rm b}$ are coefficients which can be chosen to optimize simulation efficiency. By these source terms, we can generate world-line configurations with multiple worms in both of the two sectors, A and B, at the same time. However, updating procedures using these configurations are generally complicated. Instead, we often use configurations which has just one worm for updates. Let us call a worm that works on the A(B)-bosons {\it an A}({\it B}){\it -worm}. When we apply the conventional method to two-component systems, one of A-worms or B-worms is created somewhere in the space-time stochastically. Then one of the discontinuous points({\it head}) moves stochastically along the world-lines, updating particle numbers. When the head meets the other discontinuous point({\it tail}) again, the worm is annihilated. In this algorithm, we can compute the two-point correlation functions $\langle a_{i}^{\dagger} a_{j} \rangle$ by counting the number of times the A-worm's head passes the position $(\mbox{\boldmath $r$}, \tau)=(\mbox{\boldmath $r$}_{i} - \mbox{\boldmath $r$}_{j}, 0)$ relative to the tail\cite{kawashima2004, kato2007}, where $\mbox{\boldmath $r$}_{i}$ is the coordination vector of the site $i$.

	In order to measure the pair-correlation function $\langle a_{i} b_{i}^{\dagger} a_{j}^{\dagger} b_{j} \rangle$, we introduce a new worm for updates. The worm is represented by the source term $-\eta_{\rm pair} Q_{\rm pair}$, where $Q_{\rm pair}$ is written by
	\begin{eqnarray}
		 Q_{\rm pair}=\sum_{i} \int_{0}^{\beta} d \tau \{ a_{i}^{\dagger} (\tau) b_{i} (\tau) + a_{i}(\tau) b_{i}^{\dagger}(\tau) \}.
	\end{eqnarray}
	We call this kind of worm {\it a pair-worm}. In worm creations, one of the three kinds of worms(A-worm, B-worm and pair-worm) is chosen with nearly equal probabilities, by setting $\eta_{\rm a} = \eta_{\rm b} = \eta_{\rm pair}$. The other procedures are the same as the conventional ones, including the way to measure the pair-correlation function.  It should be emphasized that creations of a conventional worm are still necessary to satisfies the ergodicity. We performed extensive checks of the code against exact diagonalization on small lattices. By using the pair-worm, simulation efficiency in the SCF phase can also be improved as explained in the latter part of this letter.


	To investigate properties of the SCF state, the equal-time correlation functions
	\begin{eqnarray}
			C_{\rm a} (r) & = & \langle a_{i}^{\dagger} a_{j} \rangle \\
			C_{\rm b} (r) & = & \langle b_{i}^{\dagger} b_{j} \rangle \\
			C_{\rm pair} (r) & = & \langle a_{i} b_{i}^{\dagger} a_{j}^{\dagger} b_{j} \rangle
	\end{eqnarray}
	are observed. In Fig. \ref{fig:correl}, we plot these correlation functions as a function of distance $r=|\mbox{\boldmath $r$}_{i} - \mbox{\boldmath $r$}_{j}|$, at various temperatures in the case of $t_{\rm a}=t_{\rm b}, 2zt_{\rm a}/U=0.6$. In the square lattice, the coordination number is $z=4$. To obtain the results of $L=64$, each run contains $5 \times 10^{5}$ Monte Carlo steps for measuring physical quantities. We performed 256 independent runs for estimating the statistical errors. The $t_{\rm a}=t_{\rm b}$ case is simulated, because the SCF state stabilizes most and is easy to analyze. Since $C_{\rm b}(r)$ is trivially equal to $C_{\rm a}(r)$ in this case, it is omitted here. At high temperatures, {\it e.g.} $T/t_{\rm a}=0.16, 0.20$, all correlation functions decays exponentially. When the temperature becomes lower($T/t_{\rm a}=0.04, 0.095$), $C_{\rm pair} (r)$ behaves as power law decay. On the other hand, $C_{\rm a}(r)$ still decays exponentially. This is a clear evidence of the SCF phase where A-particles and B-holes cannot support supercurrent individually but must form pairs to do so. 
	
\begin{figure}[t]
		\includegraphics[width=7.9cm]{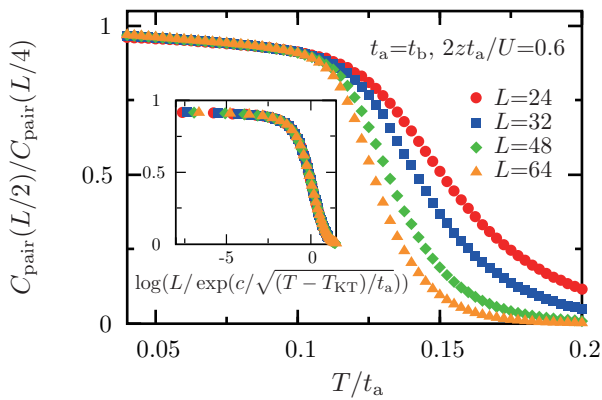}
		\caption{\label{fig:ratio_correl_pair} (Color online) Temperature dependence of the correlation ratio $C_{\rm pair}(L/2)/C_{\rm pair}(L/4)$ for different system sizes. The inset is the finite-size scaling plot. }
	\end{figure}

	The power-law decay of the correlation function is the characteristic property of the KT phase\cite{kosterlitz1973}. We can make it clearer by further analysis of $C_{\rm pair} (r)$. Let's consider the correlation ratio defined by $C_{\rm pair}(L/2)/C_{\rm pair}(L/4)$. This quantity is independent of the system size at and below the critical point like the Binder ratio. Fig. \ref{fig:ratio_correl_pair} shows the temperature dependence of the correlation ratio. It can be seen that the correlation ratio for different system sizes overlap at low temperatures. This is a sign of the KT transition. Then, we make a finite-size scaling analysis, using the scaling form for the KT transition $C_{\rm pair}( L/2)/C_{\rm pair}(L/4) (T)= f(L/\exp(c/\sqrt{(T-T_{\rm KT}) / t_{\rm a}}))$\cite{kosterlitz1974, tomita2009}, where the critical temperature $T_{\rm KT}$ and the unknown value $c$ is to be estimated. In the KT transition, the finite-size scaling analysis is generally difficult due to the the singular divergence and the logarithmic correction. However, the correlation ratio is known to be a good estimator for determining the critical point, because the logarithmic correction is cancelled\cite{tomita2002}. In the inset of Fig. \ref{fig:ratio_correl_pair}, we plot the resulting finite-size scaling of the correlation ratio. From this analysis, the estimates $T_{\rm KT}/t_{\rm a}=0.092(2)$ and $c=0.80(2)$ are obtained. 
	
	\begin{figure}[t]
		\includegraphics[width=7.3cm]{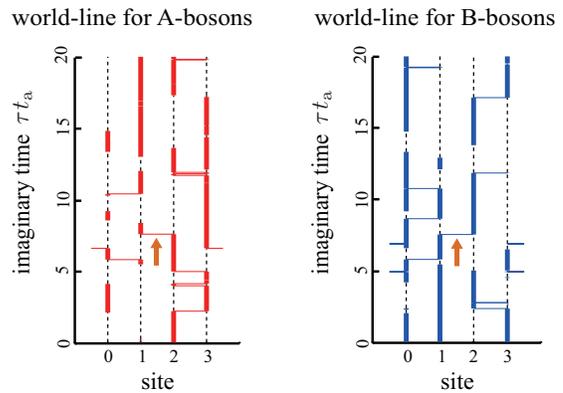}
		\caption{\label{fig:snapshot} (Color online) Cross-sectional snapshot of world-line configurations in the SCF phase. The left and the right panels are world-line configurations for the A-bosons and B-bosons respectively. Parameters are $t_{\rm a}=t_{\rm b}$, $U/t_{\rm a}=15$, $\beta t_{\rm a}=20$ and $N=4\times 4$. Solid and dashed lines stand for occupied and vacant sites respectively. The discontinuous points are the positions where bosons hop in the direction vertical to the paper. Arrows show one of the approximate pair-hoppings.}
	\end{figure}

	By snapshots of the world-lines(Fig. \ref{fig:snapshot}), we also obtain the physical picture of the SCF state as follows: Due to the strong repulsive interaction, each site is occupied by one boson(A-boson or B-boson). This is a feature of Mott states. In order to gain the hopping energy, A-boson and B-boson often hop to each other's site in bonds where an A-boson is occupied on one side and a B-boson on the other side. This pair-hopping is the origin of the pair-condensation. As a results, the net currents of A-bosons and B-bosons are in opposite directions.

	\begin{figure}[t]
		\includegraphics[width=7.9cm]{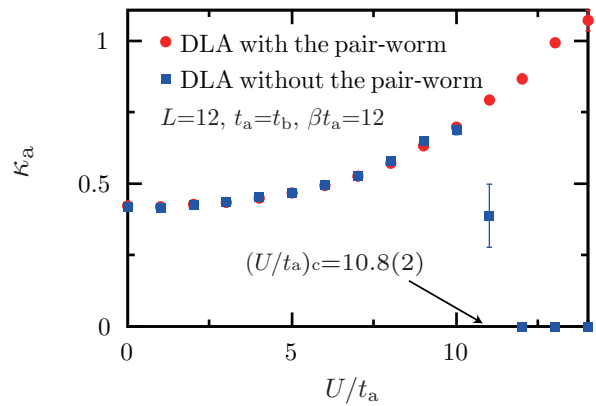}
		\caption{\label{fig:compress_a} (Color online) Comparison between the DLA without using pair-worms and the DLA which includes pair-worms by $\kappa _{\rm a}$ as a function of $U$.}
	\end{figure}
	
	We also discuss simulation efficiency of our method. For this purpose, we show the comparison between the DLA which includes the pair-worm and the DLA which does {\it not} include the pair-worm in Fig. \ref{fig:compress_a}. For the comparison, the compressibility of A-bosons
	\begin{eqnarray}
		\kappa_{\rm a} & = & \frac{1}{\langle n_{\rm a} \rangle^2} \frac{\partial \langle n_{\rm a} \rangle}{\partial \mu_{\rm a} }  
	\end{eqnarray}
	is used, because this quantity showed the most remarkable difference of all measured quantities, as explained below. In the above equation, $\mu_{\rm a}$ means the chemical potential of A-bosons and $\kappa_{\rm a}$ is proportional to the fluctuation of the total particle number of A-bosons. In both simulations, we performed $1.8 \times 10^{4}$ Monte Carlo runs for both thermalization and measuring quantities. By the finite-size-scaling analysis based on the ($d+1$) dimensional $U(1)$ universality class\cite{soyler2009}, we obtained the quantum critical point $(U/t_{\rm a})_{\rm c}=10.8(2)$. As the interspecies interaction $U$ is increased beyond this point, the system transits from two-species superfluid (2SF) phase to the SCF phase. We have observed that results by two methods in Fig. 2 are consistent within error bars in the 2SF phase, but not in the SCF phase. 
	
	To understand this discrepancy, the important fact is that the typical length of worm's movement correctly reflects the correlation length in the DLA\cite{kawashima2004}. In the SCF phase, updates by the conventional A(B)-worms are often very local, {\it i.e.} restricted in a very narrow region, because the correlation length of A(B)-bosons are short-ranged. On the other hand,  the method with the pair-worm enables to update configurations globally even in the SCF phase, because the correlation length of $a^{\dagger}b$ is divergent. When we measure the fluctuation of the total number of A(B)-bosons, global worm updates are necessary. This is because the total particle number can be changed only when the worm head goes across the periodic boundary of the imaginary axis and return to the worm tail. That is the reason why the compressibility $\kappa_{\rm a}$ incorrectly vanishes in the method without the pair-worm and does not vanish in the method including the pair-worm. Other quantities such as stiffnesses make no difference between the two methods within error bars. However, the above result means that the method without the pair-worm may produce erroneous results in the SCF phase and needs longer runs. We must be careful when we are using the conventional method, because most quantities are close to correct equilibrium values as long as we do not pay attention to the compressibility of A(B)-bosons.
	
	In conclusion, we have proposed the simple QMC technique for measuring the pair-correlation function. By this method, we have calculated several correlation functions including the pair-correlation function in order to confirm that the pair-condensation is characteristic and essential to the SCF phase. Furthermore, we demonstrated that the new worm also improves the simulation efficiency in the SCF phase. In this improvement of simulation efficiency, the important fact is that we have introduced the worm corresponding to the order parameter $a^{\dagger} b$ whose correlation length is divergent at the criticality. Our results suggest that suitable worms should be used according to what kind of correlation exits. This idea is expected to be applied to general multi-degree-of-freedom systems where various type of order parameters exit.

	The authors are grateful to Y. Kato, T. Suzuki, and Y. Tomita for valuable discussions. The present work is financially supported by the MEXT Global COE Program "the Physical Science Frontier", the MEXT Grant-in-Aid for Scientific Research (B) (22340111), for Scientific Research on Priority Areas gNovel States of Matter Induced by Frustrationh (19052004), and the  Next Generation Supercomputing Project, Nanoscience Program, MEXT, Japan.
The simulations were performed on computers at the Supercomputer Center, Institute for Solid State Physics, University of Tokyo.


\end{document}